# Talking Condition Recognition in Stressful and Emotional Talking Environments Based on CSPHMM2s

Ismail Shahin[1]

Mohammed Nasser Ba-Hutair[2]

Department of Electrical and Computer Engineering

University of Sharjah

P. O. Box 27272

Sharjah, United Arab Emirates

Tel: (971) 6 5050967

Fax: (971) 6 5050877

[1]E-mail: ismail@sharjah.ac.ae,

[2]E-mail: m2008m1033m@gmail.com

# Abstract

This work is aimed at exploiting Second-Order Circular Suprasegmental Hidden Markov Models (CSPHMM2s) as classifiers to enhance talking condition recognition in stressful and emotional talking environments (completely two separate environments). The stressful talking environment that has been used in this work uses Speech Under Simulated and Actual Stress (SUSAS) database, while the emotional talking environment uses Emotional Prosody Speech and Transcripts (EPST) database. The achieved results of this work using Mel-Frequency Cepstral Coefficients (MFCCs) demonstrate that CSPHMM2s outperform each of Hidden Markov Models (HMMs), Second-Order Circular Hidden Markov Models (CHMM2s), and Suprasegmental Hidden Markov Models (SPHMMs) in enhancing talking condition recognition in the stressful and emotional talking environments. The results also show that the performance of talking condition recognition in stressful talking environments leads that in emotional talking environments by 3.67% based on CSPHMM2s. Our results obtained in subjective evaluation by human judges fall within 2.14% and 3.08% of those obtained, respectively, in stressful and emotional talking environments based on CSPHMM2s.

# 1. Introduction

Stressful talking environments are defined as the talking environments where speakers utter their speech under the influence of stressful talking conditions such as shouted, slow, and fast talking conditions. There are many factors that create stress into the speech production process such as noisy background, emergency conditions such as that in aircraft pilot communications, high workload stress, physical environmental factors, multitasking, and physical G-force movement such as fighter cockpit pilot [1]. There are many applications of talking condition recognition in stressful talking environments. Such applications include emergency telephone message sorting, telephone banking, hospitals which include computerized stress categorization and evaluation techniques, and military voice communication applications.

Emotional talking environments are defined as the talking environments where speakers utter their speech under the effect of emotional states such as anger, happiness, and sadness. The applications of emotion recognition appear in telecommunications, human robotic interfaces, smart call centers, and intelligent spoken tutoring systems. In telecommunications, emotion recognition can be used to evaluate a caller's emotional status for telephone response services. Emotion recognition can also be used in human robotic interfaces where robots can be taught to interact with humans and identify human emotions. More applications of emotion recognition from speech can be seen in smart call centers where possible problems appearing from disappointing interactions can be detected by emotion recognition systems. Emotion recognition can be exploited in intelligent spoken tutoring systems to perceive and adjust to students' emotions when students reached a boring state during tutoring sessions [2], [3], [4].

## 2. Motivation and Literature Review

The field of stressful talking condition recognition has been studied in many occasions [1], [5], [6], [7]. Some talking conditions are designed to imitate speech under real stressful talking conditions. Bou-Ghazale and Hansen [1] and Zhou *et al*. [7] recorded and used Speech Under Simulated and Actual Stress (SUSAS) database in which eight talking conditions are used to mimic speech generated under real stressful talking conditions. These conditions are neutral, loud, soft, angry, fast, slow, clear, and question. Shahin [5] used circular hidden Markov models (CHMMs) to study talking condition identification. He used neutral, shouted, loud, slow, and fast talking conditions. Chen [6] studied talker-stress-induced intraword variability and an algorithm that pays off for the systematic changes observed based on hidden Markov models (HMMs) trained by speech tokens in different talking conditions. He used six talking conditions to simulate speech under real stressful talking conditions. The talking conditions are neutral, fast, loud, Lombard, soft, and shouted.

There are many studies that focus on the field of emotion recognition. Fragopanagos and Taylor [4] outlined their developed approach to construct an emotion-recognizing system. It is based on guidance from psychological studies of emotion, as well as from the nature of emotion in its interaction with attention. They used a Neural Network architecture to handle the fusion of different modalities. Lee and Narayanan [8] focused in one of their works on recognizing emotions from spoken language. They used a mixture of three sources of information for emotion recognition. The three sources are acoustic, lexical, and discourse. Morrison *et al*. [9] endeavored in one of their studies to improve the emotional speech classification methods based on ensemble or multi-classifier system (MCS) approaches. They also aimed to examine the differences in recognizing emotions in human speech that are obtained from different methods of acquisition. Nwe *et al*. [10] proposed in one of their works a text-independent method of emotion classification of speech based on HMMs. Casale *et al*. [11] suggested a new feature vector that contributes in improving the classification performance of emotional/stressful states of humans. The components of such a feature vector are attained from a feature subset selection method based on genetic algorithm.

In one of his prior studies [12], Shahin focused on studying and enhancing text-independent and speaker-independent talking condition identification in stressful and emotional talking environments (completely two separate environments) based on three separate and distinct classifiers. The three classifiers are HMMs, Second-Order Circular Hidden Markov Models (CHMM2s), and Suprasegmental Hidden Markov Models (SPHMMs). He concluded in that study that SPHMMs outperform each of HMMs and CHMM2s for talking condition identification in the two talking environments [12]. In the current work, the main contribution is directed towards enhancing text-independent and speaker-independent talking condition identification in each of stressful and emotional talking environments based on exploiting Second-Order Circular Suprasegmental Hidden Markov Models (CSPHMM2s) as classifiers. This work is a continuation to the work of [12]. Specifically, the main aim of the present work is to further improve talking condition recognition in these two separate talking environments based on a combination of HMMs, CHMM2s, and SPHMMs. This combination is called CSPHMM2s. In addition, one of the main objectives of this work is to discriminate between stressful talking environments and emotional talking environments based on CSPHMM2s.

Two well-known speech databases have been used in this work to test CSPHMM2s for talking condition recognition in stressful and emotional talking environments. The first database is called SUSAS database which was recorded in neutral and stressful talking environments [13], while the second one is called Emotional Prosody Speech and Transcripts (EPST) database which was collected in neutral and emotional talking environments [14].

The rest of the paper is organized as follows: The details of CSPHMM2s are given in Section 3. The speech databases used in the current work and extraction of features are explained in Section 4. Section 5 discusses the algorithm of stressful/emotional talking condition identification based on CSPHMM2s. Section 6 discusses the results obtained in this work and the experiments. Finally, Section 7 concludes this work with some remarks.

### 3. Second-Order Circular Suprasegmental Hidden Markov Models

In literature, there are many techniques, algorithms, and classifiers that have been used to classify the stressful/emotional state of a speaker through speech. HMMs have been used by: Bou-Ghazale and Hansen [1] in stressful talking environments, Nwe *et al.* [10] in emotional talking environments, and Shahin [12] in stressful and emotional talking environments. Neural Networks (NNs) have been applied by Hansen and Womack [15] in stressful talking environments and by Park and Sim [16] in emotional talking environments. Genetic Algorithms (GAs) have been implemented by Casale *et al.* [11] in stressful talking environments. Support Vector Machines (SVMs) have been exploited in emotional talking environments by Oudeyer [17] and by Kwon *et al.* [18]. In one of his works, Shahin [12] used each of HMMs, CHMM2s, and SPHMMs as classifiers in stressful and emotional talking environments.

CSPHMM2s have been developed, implemented, and evaluated by Shahin [19] to improve speaker identification performance in shouted talking environments. These models have been derived from both acoustic Second-Order Circular Hidden Markov Models and Suprasegmental Hidden markov Models. CHMM2s have been proposed, applied, and tested by Shahin to enhance speaker identification performance in emotional [20] and shouted [21] talking environments. SPHMMs have been developed, used, and assessed by Shahin for speaker recognition in emotional [20] and shouted [22] talking environments. SPHMMs have the ability to summarize several conventional HMM states into what is called suprasegmental state. Suprasegmental state has the ability to look at the observation sequence through a larger window. Such a state allows observations at rates suitable for the situation of modeling. Prosodic events at the levels of phone, syllable, word, and utterance are modeled using suprasegmental states, while acoustic events are modeled using conventional states. More information about SPHMMs can be obtained from *Ref.* [22].

Acoustic and prosodic information within CHMM2s can be combined and integrated as given by the following formula [23],

$$\log P\left(\lambda^v_{CHMM2s}, \Psi^v_{CSPHMM2s} | O\right) = (1-\alpha) \cdot \log P\left(\lambda^v_{CHMM2s} | O\right) \\ + \alpha \cdot \log P\left(\Psi^v_{CSPHMM2s} | O\right) \quad (1)$$

where $\alpha$ is a weighting factor. When:

$$\begin{cases} 0.5 > \alpha > 0 & \text{biased towards acoustic model} \\ 1 > \alpha > 0.5 & \text{biased towards prosodic model} \\ \alpha = 0 & \text{biased completely towards acoustic model and no effect of prosodic model} \\ \alpha = 0.5 & \text{not biased towards any model} \\ \alpha = 1 & \text{biased completely towards prosodic model and no impact of acoustic model} \end{cases} \quad (2)$$

$\lambda^v_{CHMM2s}$ is the acoustic second-order circular hidden Markov model of the $v^{th}$ stressful/emotional talking condition, $\Psi^v_{CSPHMM2s}$ is the suprasegmental second-order circular hidden Markov model of the $v^{th}$ stressful/emotional talking condition, $O$ is the observation vector or sequence of an utterance, $P\left(\lambda^v_{CHMM2s} | O\right)$ is the probability of the $v^{th}$ CHMM2s stressful/emotional talking condition model given the observation vector $O$, and $P\left(\Psi^v_{CSPHMM2s} | O\right)$ is the probability of the $v^{th}$ CSPHMM2s stressful/emotional talking condition model given the observation vector $O$.

The initial components of the parameters in the training phase of CHMM2s have been selected to be [21],

$$v_k(i) = \frac{1}{N} \qquad N \geq i, k \geq 1 \qquad (3)$$

where $v_k(i)$ is the initial element of the probability of an initial state distribution and $N$ is the number of states.

$$alpha_1(i,k) = v_k(i) b_{ki}(O_1) \qquad N \geq i, k \geq 1 \qquad (4)$$

where $alpha_1(i,k)$ is the initial element of the forward probability of generating the observation vector $O_1$ and $b_{ki}(O_1)$ is the element of the observation symbol probability of the observation vector $O_1$.

$$a_{ijk}^1 = \begin{cases} \frac{1}{3} & i = 1, j, k = 1, 2, ..., N \text{ or } N-1 \geq i \geq 2, i+1 \geq j \geq i-1, N \geq k \geq 1 \\ & \text{or } i = N, \ j, k = 1, 2, ..., N \\ 0 & \text{otherwise} \end{cases} \qquad (5)$$

where $a^1_{ijk}$ is the initial element of $a_{ijk}$ (CHMM2s state transition coefficients).

$$b_{ijk}^1 = \frac{1}{M} \qquad N \geq j, k \geq 1, M \geq i \geq 1 \qquad (6)$$

where $b^1_{ijk}$ is the initial element of CHMM2s observation symbol probability and $M$ is the number of observation symbols.

$$beta_T(j,k) = \frac{1}{N} \qquad N \geq j, k \geq 1 \qquad (7)$$

where $beta_T(j,k)$ is the initial element of the backward probability of creating the observation vector $O_T$.

$$P(O|\lambda) = \sum_{k=1}^{N} \sum_{i=1}^{N} alpha_T(i,k) \qquad (8)$$

where $P(O|\lambda)$ is the probability of the observation vector $O$ given the CHMM2s model $\lambda$. The reader can obtain more details about the second-order circular hidden Markov models from *Ref.* [21].

CSPHMM2s are superior to each of First-Order Left-to-Right Suprasegmental Hidden Markov Models (LTRSPHMM1s), Second-Order Left-to-Right Suprasegmental Hidden Markov Models (LTRSPHMM2s), and First-Order Circular Suprasegmental Hidden Markov Models (CSPHMM1s). This is because the characteristics of LTRSPHMM1s, LTRSPHMM2s, and CSPHMM1s are combined and integrated into CSPHMM2s:

1. The state sequence in second-order models is a second-order chain where the stochastic process is characterized by a 3-D matrix because the state-transition probability at time $t+1$ depends on the states of the chain at times $t$ and $t$-1. On the other hand, the state sequence in first-order models is a first-order chain where the stochastic process is characterized by a 2-D matrix since the state-transition probability at time $t+1$ depends only on the state at time $t$. Thus, the stochastic process that is specified by a 3-D matrix yields higher talking condition identification performance than that specified by a 2-D matrix.

2. Markov chain in circular models is more powerful and more efficient than that possessed by left-to-right models to model the changing statistical characteristics that exist in the actual observations of speech signals. The absorbing state in the left-to-right models rules the fact that the remaining of a single observation sequence provides no additional information about earlier states once the underlying Markov chain reaches the absorbing state. In speaker identification, it is true that a Markov chain should be able to revisit the earlier states since the states of HMMs reflect the vocal organic configuration of the speaker. Therefore, the vocal organic configuration of the speaker is reflected to states more properly using circular models than that using left-to-right models. Consequently, it is improper to employ left-to-right models having one absorbing state for speaker identification.

Fig. 1 demonstrates an example of a fundamental structure of CSPHMM2s that has been obtained from CHMM2s. This figure is made up of six second-order acoustic hidden Markov states: $q_1, q_2,..., q_6$ located in a circular form. $p_1$ is a second-order suprasegmental state which consists of $q_1$, $q_2$, and $q_3$. $p_2$ is a second-order suprasegmental state which is composed of $q_4$, $q_5$, and $q_6$. The suprasegmental states $p_1$ and $p_2$ are arranged in a circular form. $p_3$ is a second-order suprasegmental state which is comprised of $p_1$ and $p_2$. $a_{ij}$ is the transition probability between the $i^{th}$ acoustic hidden Markov state and the $j^{th}$ acoustic hidden Markov state, while $b_{ij}$ is the transition probability between the $i^{th}$ suprasegmental state and the $j^{th}$ suprasegmental state. The transition matrix, **B**, of such a structure using the two suprasegmental states $p_1$ and $p_2$ can be defined using the positive coefficients $b_{ij}$ as,

$$B = \begin{bmatrix} b_{11} & b_{12} \\ b_{21} & b_{22} \end{bmatrix}$$

## 4. Speech Databases and Extraction of Features

### 4.1 Speech Under Simulated and Actual Stress (SUSAS) Database

SUSAS database is comprised of five domains, covering an ample range of stresses and emotions. The database contains both simulated speech under stress (Simulated Domain) and actual speech under stress (Actual Domain). A total of 32 speakers (19 male and 13 female), with ages spanning from 22 to 76 years were used to utter more than 16,000 utterances [13]. In the present work, only 20 different words (10 words were used for training and the rest were used for testing) uttered by 8 speakers (5 speakers were used for training and the remaining were used for testing) 2 times (2 repetitions per word) talking in 6 stressful talking conditions were used. These talking conditions are neutral, angry, slow, loud, soft, and fast.

### 4.2 Emotional Prosody Speech and Transcripts (EPST) Database

This database is made up of 8 professional speakers (3 actors and 5 actresses) uttering a series of semantically neutral utterances comprising of dates and numbers spoken in 15 different emotions [14]. In the current work, only 20 different utterances (10 utterances were used for training and the remaining were used for testing) uttered by 8 speakers (5 speakers were used for training and the rest were used for testing) talking in 6 emotions were used. The emotions are neutral, hot anger, sadness, happiness, disgust, and panic.

**4.3 Extraction of Features**

The phonetic content of speech signals in the two databases of this work was represented by Mel-Frequency Cepstral Coefficients (static MFCCs) and delta Mel-Frequency Cepstral Coefficients (delta MFCCs). These coefficients have been broadly used by many researchers in the areas of speech recognition [7], [24], [25], speaker recognition [26], [27], and stressful/emotional talking condition recognition [8], [18], [28]. In the present work, MFCC feature analysis was used to form the observation vectors for CSPHMM2s in the stressful and emotional talking environments.

A 32-dimension MFCC (16 static MFCCs and 16 delta MFCCs) feature analysis was used to form the observation vectors in CSPHMM2s. The number of conventional states in CHMM2s was 6, while the number of suprasegmental states in CSPHMM2s was 2 (each suprasegmental state was comprised of 3 conventional states). The number of mixture components, $M$, was 10 per state with a continuous mixture observation density was selected for CSPHMM2s.

**5. Stressful/Emotional Talking Condition Identification Algorithm Based on CSPHMM2s**

The training phase of CSPHMM2s in each of the SUSAS and EPST databases is similar to the training phase of conventional CHMM2s. In the training phase of CSPHMM2s, suprasegmental models are trained on top of acoustic models. In each training phase of the two databases, one reference model per stressful/emotional talking condition has been derived using 5 of the 8 speakers uttering 10 utterances with a repetition of 2 times per utterance. The total number of utterances that has been used in this phase to derive each CSPHMM2s stressful/emotional talking condition model is 100 (5 speakers × 10 utterances × 2 times/utterance). The two training phases are completely separate from each other. In the test phase of each database, each one of the 3 remaining speakers uses different 10 utterances with a repetition of 2 times per utterance under each stressful/emotional talking condition (text-independent and speaker-independent experiments). The total number of utterances that has been used in this phase per database is 360 (3 speakers × 10 utterances × 2 times/utterance × 6 stressful/emotional talking conditions). The probability of generating every utterance is computed based on CSPHMM2s as given in the following formula,

$$E^* = \arg\max_{6 \geq e \geq 1} \left\{ P\left(O \mid \lambda^e_{\text{CHMM2s}}, \Psi^e_{\text{CSPHMM2s}}\right) \right\} \qquad (9)$$

where, $E^*$ is the index of the identified stressful/emotional talking condition, $O$ is the observation vector that corresponds to the unknown stressful/emotional talking condition, and $P\left(O \mid \lambda^e_{CHMM2s}, \Psi^e_{CSPHMM2s}\right)$ is the probability of the observation sequence $O$ given the $e^{\text{th}}$ CSPHMM2s stressful/emotional talking condition model ($\lambda^e$, $\Psi^e$). A block diagram of stressful/emotional talking condition recognizer based on CSPHMM2s is shown in Fig. 2.

## 6. Results and Discussion

In the current work, CSPHMM2s have been exploited as classifiers to enhance talking condition recognition in each of stressful and emotional talking environments. In such classifiers, the value of the weighting factor ($\alpha$) has been selected to be equal to 0.5 to avoid biasing towards any model.

Talking condition identification performance in stressful talking environments based on HMMs, CHMM2s, SPHMMs, and CSPHMM2s using SUSAS database is given in Table 1. This table yields an average stressful talking condition identification performance of 64.4%, 68.5%, 72.4%, and 76.3% based on HMMs, CHMM2s, SPHMMs, and CSPHMM2s, respectively.

A statistical significance test has been carried out to investigate whether stressful/emotional talking condition identification performance differences (stressful/emotional talking condition identification performance based on CSPHMM2s and that based on each of HMMs, CHMM2s, and SPHMMs) are real or simply due to statistical fluctuations. The statistical significance test has been performed based on the Student's *t* distribution test as given by the following formula,

$$t_{model\,x, model\,y} = \frac{\overline{x}_{model\,x} - \overline{x}_{model\,y}}{SD_{pooled}} \qquad (10)$$

where,

$\overline{x}_{model\,x}$: is the mean of the first sample (model *x*) of size *n*.

$\overline{x}_{model\,y}$: is the mean of the second sample (model *y*) of the same size.

SD $_{pooled}$: is the pooled standard deviation of the two samples (models *x* and *y*) given as,

$$SD_{pooled} = \sqrt{\frac{SD^2_{model\,x} + SD^2_{model\,y}}{2}} \quad (11)$$

where,

SD $_{model\ x}$: is an estimate of the standard deviation of the average of the first sample (model *x*) of size *n*.

SD $_{model\ y}$: is an estimate of the standard deviation of the average of the second sample (model *y*) of the same size.

Based on Table 1, the calculated *t* value between CSPHMM2s and each of HMMs, CHMM2s, and SPHMMs using SUSAS database is given in Table 2. This table evidently shows that every calculated *t* value is greater than the tabulated critical value $t_{0.05}$ = 1.645 at 0.05 significant level. Therefore, it is apparent from Table 2 that CSPHMM2s are superior to HMMs, CHMM2s, and SPHMMs for stressful talking condition identification; *i.e.* the difference is significant and it is not due to a random error.

Another way of seeing this is by constructing a confidence interval for the actual difference of two means of two specific models. For example, if $\mu_x$ is the mean of model *x* (say CSPHMM2s) and $\mu_y$ is the mean of model *y* (say HMMs), the 95% confidence interval of $\mu_x - \mu_y$ is $(\bar{x} - \bar{y}) \pm 1.645\,(SD_{pooled}) = (76.3 - 64.4)$ $\pm 1.645\,(6.201) = [1.699,\ 22.101]$. Since all values in the interval are positive, there is a significant positive difference between the means of the two models. In other words, we are 90% confident that $\mu_x - \mu_y$ is between 1.699 and 22.101. The confidence intervals between CSPHMM2s and each of HMMs, CHMM2s, and SPHMMs using SUSAS database are calculated in Table 2.

Fig. 3 illustrates a relative improvement percentage per each stressful talking condition of using CSPHMM2s over each of HMMs, CHMM2s, and SPHMMs when α = 0.5. It is apparent from this figure that the maximum relative improvement percentage takes place under the slow talking condition (24.0%), while the minimum relative improvement percentage happens under the neutral talking condition (2.6%).

Table 3 demonstrates a confusion matrix which represents percentage of confusion of a test stressful talking condition of SUSAS database with the other stressful talking conditions based on CSPHMM2s when α = 0.5. This table demonstrates the following:

a) The most easily recognizable stressful talking condition is neutral (97%). Consequently, the highest talking condition identification performance in stressful talking environments is neutral.

b) The least easily recognizable stressful talking condition is angry (63.5%). Thus, the least talking condition identification performance in such talking environments is angry.

c) The last column ('Fast' column), for example, shows that 11% of the utterances that were portrayed in a fast talking condition were evaluated as uttered in an angry talking condition, 4% of the utterances that were produced in a fast talking condition were identified as generated in a slow talking condition. This column shows that fast talking condition has the highest confusion percentage with angry talking condition (11%). Therefore, fast talking condition is highly confusable with angry talking condition. This column also illustrates that fast talking condition has the least confusion percentage with neutral talking condition (0%). Thus, fast talking condition is not confusable at all with neutral talking condition. This column says that 73.5% (in bold) of the utterances that were uttered in a fast talking condition were identified correctly.

Emotion identification performance based on HMMs, CHMM2s, SPHMMs, and CSPHMM2s using EPST database is given in Table 4. This table gives average emotion identification performance of 63.0%, 67.4%, 70.5%, and 73.6% based on HMMs, CHMM2s, SPHMMs, and CSPHMM2s, respectively. The calculated $t$ value between CSPHMM2s and each of HMMs, CHMM2s, and SPHMMs based on Table 4 is given in Table 5. This table clearly shows that every calculated $t$ value is higher than the tabulated critical value $t_{0.05} = 1.645$. Hence, it is evident from Table 5 that CSPHMM2s outperform each of HMMs, CHMM2s, and SPHMMs in emotion identification. The confidence intervals between CSPHMM2s and each of HMMs, CHMM2s, and SPHMMs using EPST database are computed in Table 5.

Fig. 4 shows a relative improvement percentage per each emotion of using CSPHMM2s over each of HMMs, CHMM2s, and SPHMMs. This figure clearly shows that the highest relative improvement percentage occurs under the hot anger emotion (32.2%), while the least relative improvement percentage happens under the neutral emotion (1.6%). A confusion matrix that yields a percentage of confusion of a test emotion with the other emotions using EPST database based on CSPHMM2s when $\alpha = 0.5$ is given in Table 6.

Comparing CSPHMM2s with each of HMMs, CHMM2s, and SPHMMs in each talking environment, it is evident that CSPHMM2s outperform HMMs, CHMM2s, and SPHMMs in each of the stressful and emotional talking environments. This may be accredited to the fact that the characteristics of HMMs, CHMM2s, and SPHMMs are all combined and integrated into the characteristics of CSPHMM2s.

CSPHMM2s have been compared with LTRSPHMM1s, LTRSPHMM2s, and CSPHMM1s in each of the stressful and emotional talking environments. The average talking condition identification performance in each talking environment based on these four classifiers when the value of the weighting factor is equal to 0.5 is shown in Fig. 5. The calculated $t$ value between CSPHMM2s and each of LTRSPHMM1s, LTRSPHMM2s, and CSPHMM1s is given in Table 7. This table apparently shows that every calculated $t$ value is larger than the tabulated critical value $t_{0.05} = 1.645$. Consequently, CSPHMM2s are superior to the other three classifiers in each of stressful and emotional talking environments. The confidence intervals between CSPHMM2s and each of LTRSPHMM1s, LTRSPHMM2s, and CSPHMM1s using SUSAS and EPST databases are calculated in this table.

Based on CSPHMM2s and using the achieved results of Table 1 and Table 4, the calculated $t$ value between SUSAS database and EPST database is $t_{\text{SUSAS, EPST}} = 1.699$ which is higher than the tabulated critical value $t_{0.05} = 1.645$. Therefore, there is a significant difference between stressful talking condition identification performance and emotional talking condition identification performance based on such classifiers. Using these two tables, the average stressful talking condition identification performance based on HMMs, CHMM2s, SPHMMs, and CSPHMM2s is higher than the average emotion identification performance by a percentage of 2.22%, 1.63%, 2.70%, and 3.67%, respectively. Therefore, it is evident that CSPHMM2s are more efficient classifiers than the other three classifiers in discriminating between stressful and emotional talking conditions.

The achieved results in the current work of talking condition identification performance in each of stressful and emotional talking environments are higher than those reported in prior studies:

1) Nwe *et al.* [10] attained an average percentage of classification accuracy of 59.0% using MFCCs as feature parameters and HMMs as classifiers in an emotional environment that is comprised of 6 basic emotions (anger, disgust, fear, joy, sadness, and surprise).

2) Casale *et al.* [11] reported 44.6% as an average 4-stressful talking condition identification performance of text-independent multistyle classification using MFCCs. They also obtained 66.0% as an average 4-stressful talking condition identification performance of text-independent multistyle classification using a 16-GA feature.

3) Oudeyer [17] obtained 55.6% as an average emotion identification performance based on an unsupervised series experiment for a database consisting of 5 emotions.

4) Kwon *et al.* [18] achieved an average talking condition identification performance of 70.1% based on a Gaussian SVM for a 4-class talking condition classification using SUSAS database. They also obtained, using AIBO database, an average emotion identification performance of 42.3% for a 5-class emotion identification.

To assess the attained results in the present work, four more experiments have been separately performed. The four experiments are:

i) Experiment 1: Talking condition recognition based on CSPHMM2s in each of stressful and emotional talking environments has been evaluated on two separate collected speech databases. The two databases are described as follows:

A total of 30 (15 male and 15 female students) non-professional (the database is closer to the real-life data than to the acted data) healthy adult Native American speakers were used to utter separately each of stressful and emotional speech databases. Each speaker was asked to utter 8 sentences where each sentence was uttered 9 times under each stressful talking condition (neutral, shouted, slow, loud, soft, and fast) and each emotion (neutral, angry, sad, happy, disgust, and fear). The total number of utterances uttered per talking environment was 12960 (30 speakers $\times$ 8 sentences × 9 utterances/sentence × 6 stressful/emotional talking conditions). In each database, the speakers uttered the desired sentences naturally. These speakers were allowed to hear some recorded sentences before uttering the required databases. The speakers were not allowed to practice uttering sentences under any stressful/emotional talking condition in advance. These sentences are:

1) *He works five days a week.*
2) *The sun is shining.*
3) *The weather is fair.*
4) *The students study hard.*
5) *Assistant professors are looking for promotion.*
6) *University of Sharjah.*
7) *Electrical and Computer Engineering Department.*
8) *He has two sons and two daughters.*

The two speech databases were separately captured using a speech acquisition board with a 16-bit linear coding A/D converter and sampled at a sampling rate of 16 kHz. These databases were 16-bit per sample linear data. The sampled signals were pre-emphasized and then segmented into frames of 16 ms each with 9 ms overlap between successive frames. Half of every database has been used in the training phase, while the other half of every database has been used in the test phase (text-independent and speaker-independent experiment in each database).

Table 8 and Table 9 demonstrate talking condition identification performance based on CSPHMM2s in each of stressful and emotional talking environments, respectively, using the collected databases. Table 8 yields stressful talking condition identification performance of 75.6%, while Table 9 gives emotional talking condition identification performance of 72.8%. Based on these classifiers and using the results of the two tables, the calculated $t$ value between the collected stressful database and the collected emotional database is $t_{stressful, emotional} = 1.782$ which is larger than the tabulated critical value $t_{0.05} = 1.645$. Therefore, there is a significant distinction between stressful talking condition identification performance and emotional talking condition identification performance based on CSPHMM2s.

ii) Experiment 2: The achieved results of stressful talking condition identification performance using SUSAS database and emotional talking condition identification performance using EPST database based on CSPHMM2s have been compared with those based on the state-of-the-art models and classifiers. Table 10 demonstrates average stressful talking condition identification performance using SUSAS database and average emotional talking condition identification performance using EPST database based on each of CSPHMM2s, Support Vector Machine (SVM) [29], [30], Genetic Algorithm (GA) [31], [32], and Vector Quantization (VQ) [33], [34]. This table evidently shows that CSPHMM2s outperform each of SVM, GA, and VQ for stressful and emotional talking condition identification.

In SVM, the kernel function that has been used in the training and testing phases of stressful/emotional talking condition is the Gaussian Radial

Basis Function (GRBF). Unlike the VQ model, the positive and negative distances to the hyper-planes are used. For a frame vector, the score is the maximum distance among all the distances to the hyper planes. In the identification stage, an input utterance is scored using the SVMs of each reference stressful/emotional talking condition and the distance accumulated over the entire input utterance is used to make the identification decision. The goal is to find the maximum distance from all SVMs and then compute the average distance *D* that results from an utterance [29], [30].

In GA, a well-known Simple Genetic Algorithm (SGA) has been used to search for optimal set of weights in stressful/emotional talking condition. For each candidate set of weights (*W*), a codebook *C(W)* is computed and the whole database labeled according to *W* and *C(W)*. Then, the number of labels and the number of times the label appears are counted for each training subset, and the stressful/emotional talking condition model is estimated. In SGA, chromosomes are comprised of 38 genes where each one encodes a feature weight. Eight bits have been reserved for each weight (gene values span from 0 to 255) to minimize the computational costs as much as possible. Offspring is bred by first choosing and then mixing two parents in the present population. One of the parents has been chosen based on the fitness-proportional criterion, while the second one has been selected based on the tournament method by picking the fittest of 7 arbitrarily selected individuals [31], [32], [35].

In VQ, stressful/emotional talking condition model has been derived by taking the MFCC features 2D-matrix and randomly selecting 16 frames from them. The rest of the frames are then divided into 16 groups based on their Euclidean distance from the chosen frames at the first step. Mean vector is then calculated for each group by summing the vectors together and then dividing the resulting vector by the number of frames in that group. This process is continuously repeated until it reaches to a point where the mean vectors are no longer changing. At the end of the process, there is a one stressful/emotional talking condition model for every stressful/emotional talking condition [33], [34].

iii) Experiment 3: Talking condition recognition in each of SUSAS and EPST databases has been evaluated based on CSPHMM2s for different values of the weighting factor ($\alpha$). The average talking condition identification performance for distinct values of $\alpha$ (0.0, 0.1, …, 0.9, 1.0) using SUSAS and EPST databases is illustrated in Fig. 6 and Fig. 7, respectively. The two figures demonstrate that the average talking condition identification performance (excluding the neutral talking condition) has been significantly improved as the value of the weighting factor grows. Thus, it can be concluded from this experiment that CSPHMM2s have more effect on talking condition identification performance than CHMM2s.

iv) Experiment 4: An informal subjective assessment for each of stressful talking condition identification and emotion identification (completely two separate assessments) using SUSAS and EPST databases has been carried out using 10 human listeners. These listeners are non-professional healthy adult Native American speakers. In this assessment, a total of 480 utterances per talking environment (8 speakers × 6 stressful/emotional talking conditions × 10 utterances) have been used. These listeners are asked in each evaluation to recognize the unknown stressful/emotional talking condition. The average talking condition identification performance using SUSAS and EPST databases is 74.7% and 71.4%, respectively. Based on these two averages, the average stressful talking condition identification performance is greater than the average emotion identification performance by 4.62%. Hence, stressful talking conditions can be discriminated from emotional talking conditions based on subjective assessments.

Using SUSAS database, the calculated $t$ value between the results obtained based on CSPHMM2s and those obtained based on the subjective assessment is $t_{\text{CSPHMM2s, sub. ass. (SUSAS)}} = 0.714$. Using EPST database, the calculated $t$ value between the results achieved based on CSPHMM2s and those attained based on the subjective assessment is $t_{\text{CSPHMM2s, sub. ass. (EPST)}} = 0.832$. The two calculated $t$ values are smaller than the tabulated critical value $t_{0.05} = 1.645$. Therefore, stressful/emotional talking condition identification performance based on CSPHMM2s and that based on the stressful/emotional subjective evaluation are very close.

## 7. Concluding Remarks

In this work, we focused our work on improving talking condition identification performance in each of stressful and emotional talking environments based on CSPHMM2s using global and local speech databases. Some concluding remarks can be drawn in this work. First, talking condition recognition in stressful and emotional talking environments based on CSPHMM2s outperforms that based on each of HMMs, CHMM2s, and SPHMMs. This is because CSPHMM2s possess the characteristics of the three models (HMMs, CHMM2s, and SPHMMs). Second, CSPHMM2s are superior to the state-of-the-art models and classifiers such as SVM, GA, and VQ for stressful and emotional talking condition identification. Third, it is apparent from this work that stressful talking condition identification performance is greater than emotion identification performance based on CSPHMM2s. CSPHMM2s are more capable than each of HMMs, CHMM2s, and SPHMMs to discriminate between stressful and emotional talking environments. Finally, this work clearly shows that the highest stressful/emotional talking condition identification performance takes place when CSPHMM2s are completely biased towards suprasegmental models and no impact of acoustic models.

This work has some limitations. Firstly, the number of speakers in each of SUSAS and EPST databases is limited. Secondly, CSPHMM2s do not give ideal stressful/emotional talking condition identification performance. More thorough study and investigation are planned for future work.

**Acknowledgements**

The authors wish to thank Prof. Mohammad Fraiwan Al-Saleh/ Prof. of Statistics at the Yarmouk University-Jordan for his valuable help in the statistical part of this work.

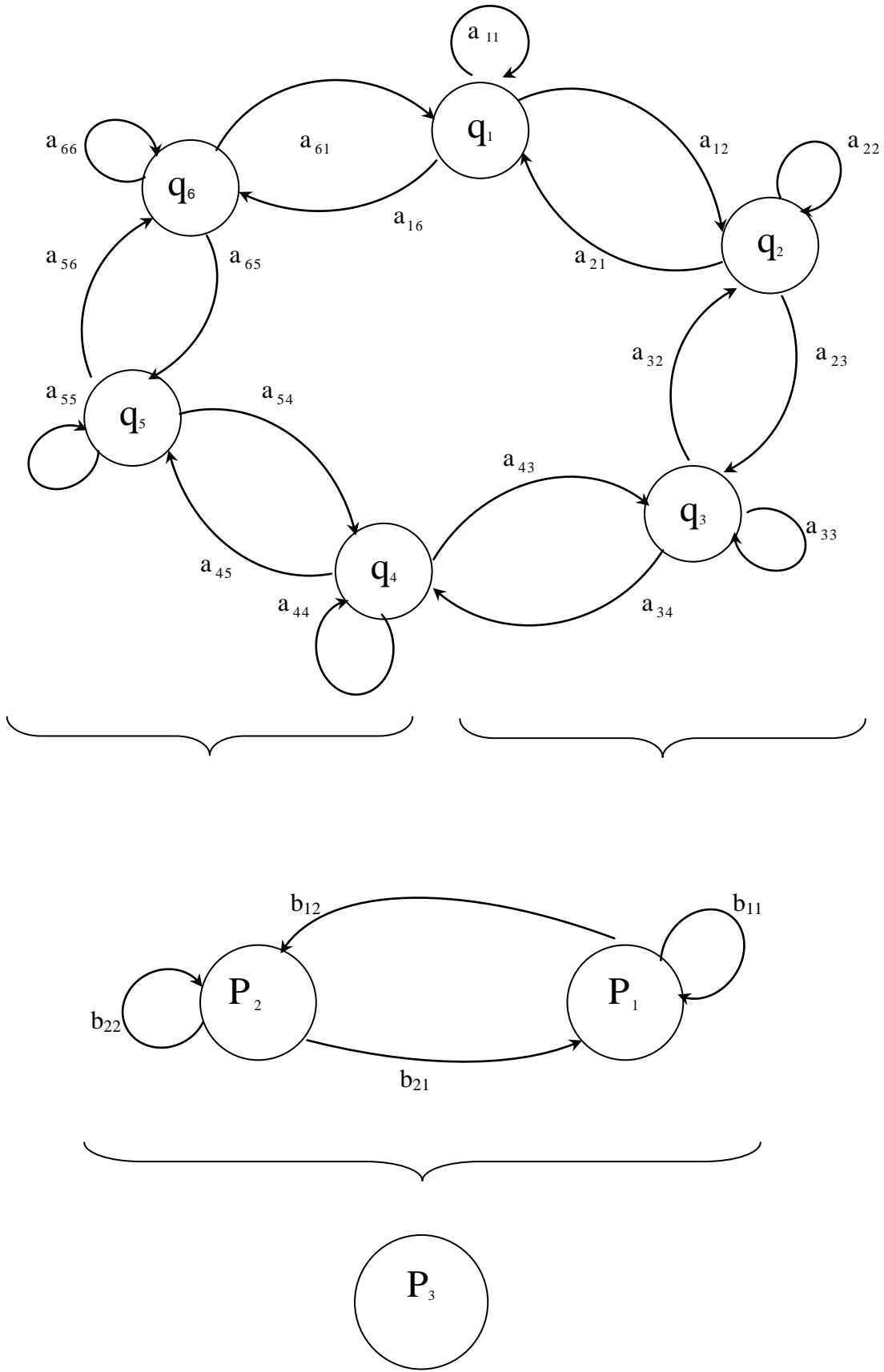

**Figure 1.** Basic structure of CSPHMM2s derived from CHMM2s

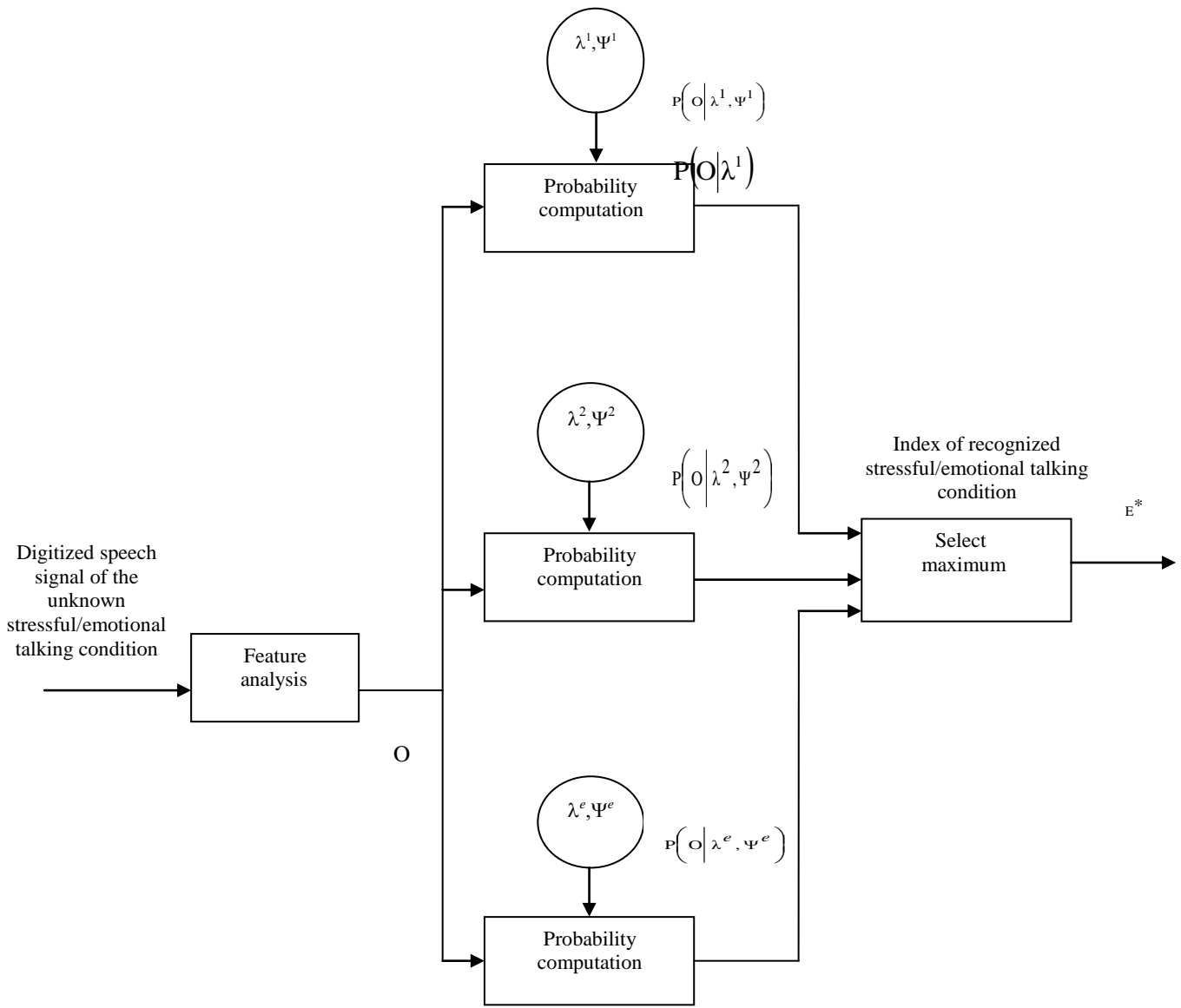

**Figure 2.** Block diagram of stressful/emotional talking condition recognizer based on CSPHMM2s

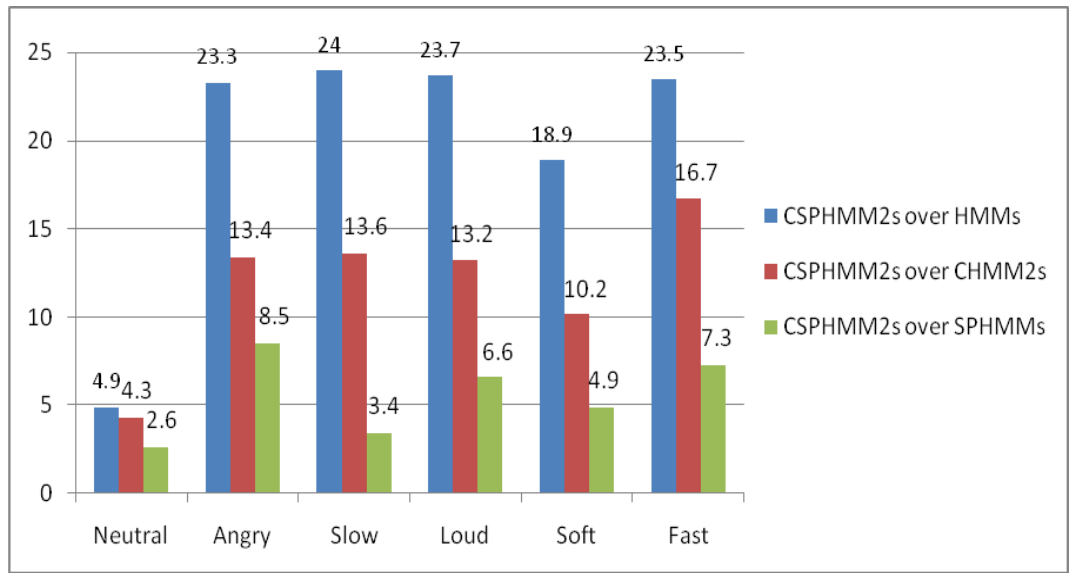

**Figure 3.** Relative improvement percentage per each stressful talking condition of using CSPHMM2s over each of HMMs, CHMM2s, and SPHMMs ($\alpha = 0.5$)

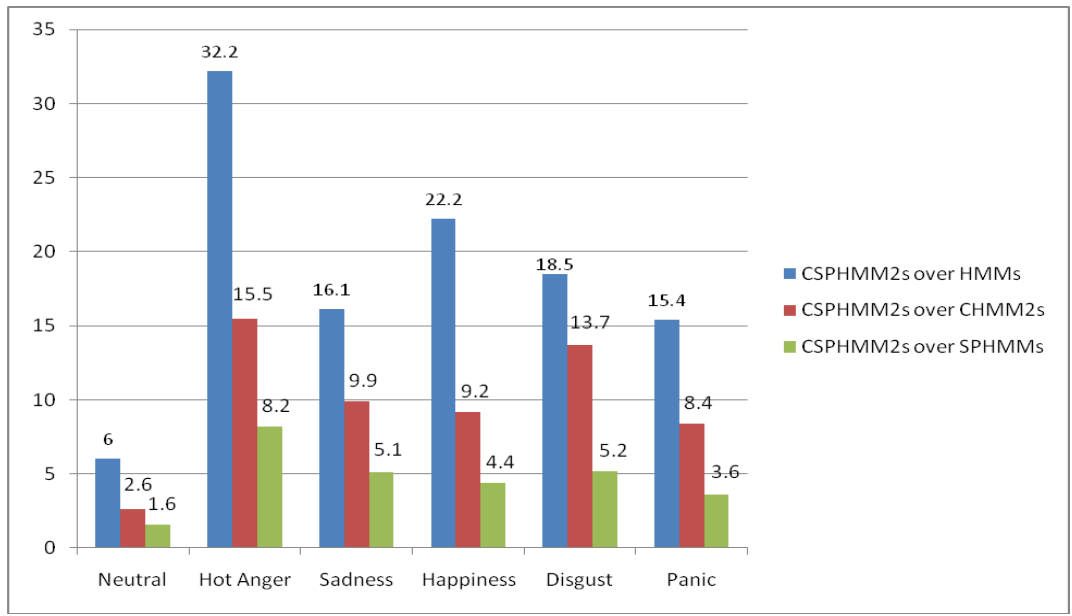

**Figure 4.** Relative improvement percentage per each emotion of using CSPHMM2s over each of HMMs, CHMM2s, and SPHMMs ($\alpha = 0.5$)

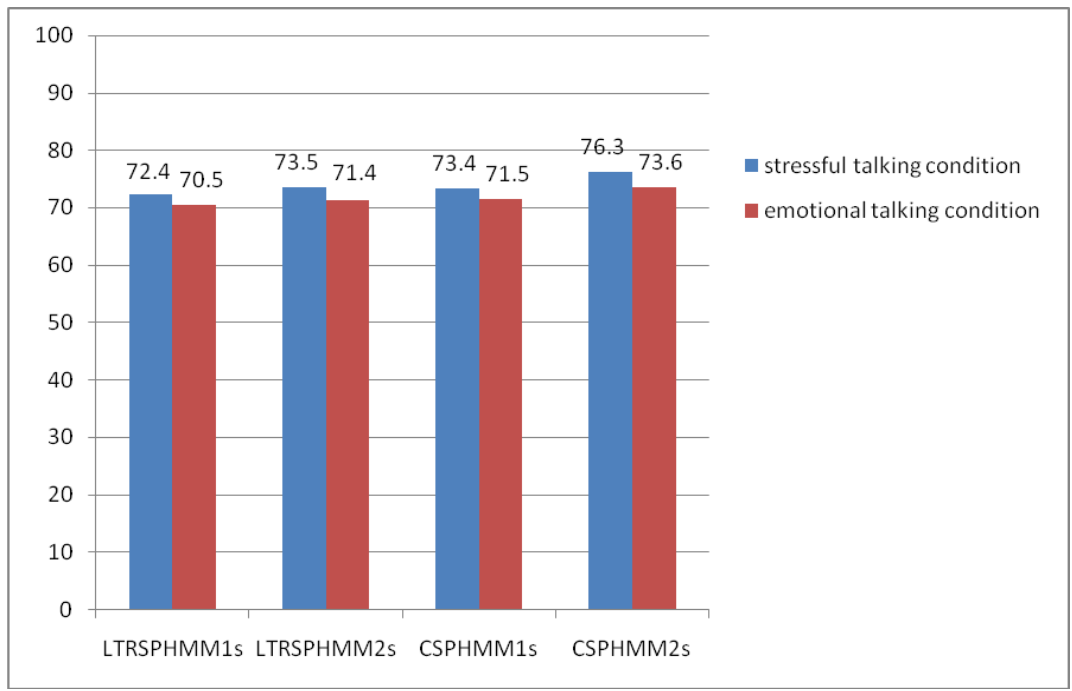

**Figure 5.** Average talking condition identification performance (%) in each of stressful and emotional talking environments based on LTRSPHMM1s, LTRSPHMM2s, CSPHMM1s, and CSPHMM2s

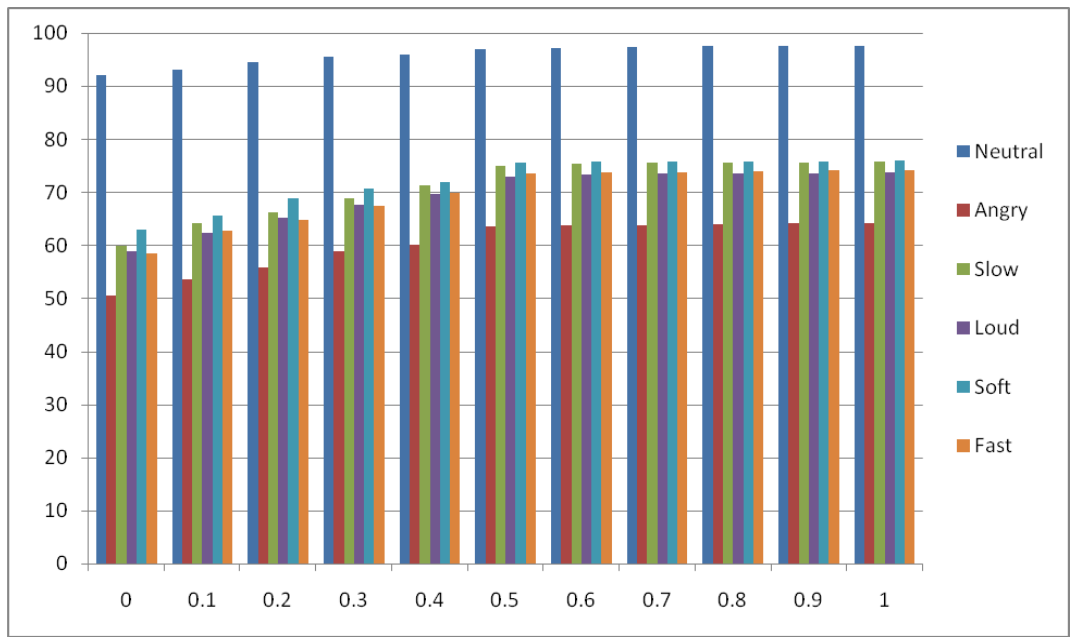

**Figure 6.** Average stressful talking condition identification performance (%) versus the weighting factor ($\alpha$) using SUSAS database based on CSPHMM2s

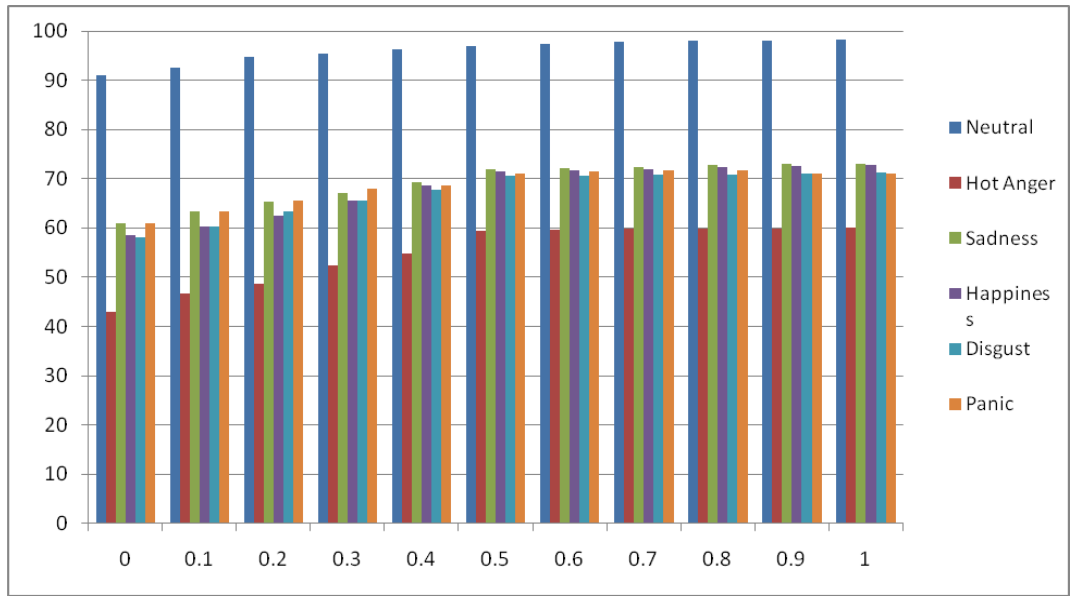

**Figure 7.** Average emotional talking condition identification performance (%) versus the weighting factor ($\alpha$) using EPST database based on CSPHMM2s

Table 1

Talking condition identification performance in stressful talking environments using SUSAS database based on HMMs, CHMM2s, SPHMMs, and CSPHMM2s when α = 0.5

| Model | Gender | Identification performance under each stressful talking condition (%) | | | | | |
|---|---|---|---|---|---|---|---|
| | | Neutral | Angry | Slow | Loud | Soft | Fast |
| HMMs | Male | 92 | 51 | 60 | 59 | 64 | 60 |
| | Female | 93 | 52 | 61 | 59 | 63 | 59 |
| | Average | 92.5 | 51.5 | 60.5 | 59 | 63.5 | 59.5 |
| CHMM2s | Male | 93 | 56 | 65 | 65 | 69 | 62 |
| | Female | 93 | 56 | 67 | 64 | 68 | 64 |
| | Average | 93 | 56 | 66 | 64.5 | 68.5 | 63 |
| SPHMMs | Male | 95 | 58 | 73 | 68 | 72 | 69 |
| | Female | 94 | 59 | 72 | 69 | 72 | 68 |
| | Average | 94.5 | 58.5 | 72.5 | 68.5 | 72 | 68.5 |
| CSPHMM2s | Male | 97 | 63 | 76 | 73 | 76 | 73 |
| | Female | 97 | 64 | 74 | 73 | 75 | 74 |
| | Average | 97 | 63.5 | 75 | 73 | 75.5 | 73.5 |

Table 2

Calculated *t* value and confidence interval between CSPHMM2s and each of HMMs, CHMM2s, and SPHMMs using SUSAS database

| $t_{model1,\ model2}$ | Calculated t value | Confidence interval |
|---|---|---|
| $t_{CSPHMM2s,\ HMMs}$ | 1.918 | [1.699, 22.101] |
| $t_{CSPHMM2s,\ CHMM2s}$ | 1.825 | [0.725, 14.875] |
| $t_{CSPHMM2s,\ SPHMMs}$ | 1.794 | [0.266, 7.534] |

Table 3

Confusion matrix in stressful talking environments using SUSAS database based on CSPHMM2s when $\alpha = 0.5$

| Talking condition | Percentage of confusion of a test stressful talking condition with the other stressful talking conditions (%) | | | | | |
|---|---|---|---|---|---|---|
| | Neutral | Angry | Slow | Loud | Soft | Fast |
| Neutral | **97** | 3 | 4 | 1 | 5 | 0 |
| Angry | 0 | **63.5** | 3 | 16 | 1 | 11 |
| Slow | 2 | 4 | **75** | 3 | 13 | 4 |
| Loud | 0 | 18 | 2 | **73** | 3 | 8 |
| Soft | 1 | 4 | 13 | 3 | **75.5** | 3.5 |
| Fast | 0 | 7.5 | 3 | 4 | 2.5 | **73.5** |

Table 4

Emotion identification performance in emotional talking environments using EPST database based on HMMs, CHMM2s, SPHMMs, and CSPHMM2s when α = 0.5

| Model | Gender | Identification performance under each emotion (%) | | | | | |
|---|---|---|---|---|---|---|---|
| | | Neutral | Hot Anger | Sadness | Happiness | Disgust | Panic |
| HMMs | Male | 91 | 45 | 62 | 58 | 60 | 62 |
| | Female | 92 | 45 | 62 | 59 | 59 | 61 |
| | Average | 91.5 | 45 | 62 | 58.5 | 59.5 | 61.5 |
| CHMM2s | Male | 95 | 52 | 66 | 65 | 62 | 65 |
| | Female | 94 | 51 | 65 | 66 | 62 | 66 |
| | Average | 94.5 | 51.5 | 65.5 | 65.5 | 62 | 65.5 |
| SPHMMs | Male | 96 | 56 | 68 | 69 | 67 | 68 |
| | Female | 95 | 54 | 69 | 68 | 67 | 69 |
| | Average | 95.5 | 55 | 68.5 | 68.5 | 67 | 68.5 |
| CSPHMM2s | Male | 97 | 59 | 72 | 71 | 70 | 71 |
| | Female | 97 | 60 | 72 | 72 | 71 | 71 |
| | Average | 97 | 59.5 | 72 | 71.5 | 70.5 | 71 |

Table 5

Calculated *t* value and confidence interval between CSPHMM2s and each of

HMMs, CHMM2s, and SPHMMs using EPST database

| $t_{model1,\ model2}$ | Calculated *t* value | Confidence interval |
|---|---|---|
| $t_{CSPHMM2s,\ HMMs}$ | 1.846 | [1.141, 20.059] |
| $t_{CSPHMM2s,\ CHMM2s}$ | 1.795 | [0.276, 12.124] |
| $t_{CSPHMM2s,\ SPHMMs}$ | 1.758 | [0.220, 5.980] |

Table 6

Confusion matrix in emotional talking environments using EPST database based on CSPHMM2s when $\alpha = 0.5$

| Talking condition | Percentage of confusion of a test emotion with the other emotions (%) | | | | | |
|---|---|---|---|---|---|---|
| | Neutral | Hot Anger | Sadness | Happiness | Disgust | Panic |
| Neutral | **97** | 1.5 | 1 | 5 | 2.5 | 2 |
| Hot Anger | 0 | **59.5** | 9 | 5 | 14 | 7 |
| Sadness | 1 | 8 | **72** | 2.5 | 5 | 9 |
| Happiness | 1 | 6 | 3 | **71.5** | 3 | 2 |
| Disgust | 0 | 17 | 5 | 8 | **70.5** | 9 |
| Panic | 1 | 8 | 10 | 8 | 5 | **71** |

Table 7

Calculated *t* value and confidence interval between CSPHMM2s and each of LTRSPHMM1s, LTRSPHMM2s, and CSPHMM1s when $\alpha = 0.5$ using SUSAS and EPST databases

| $t_{model1, model2}$ | Calculated *t* value and confidence interval | | | |
|---|---|---|---|---|
| | SUSAS | | EPST | |
| | Calculated *t* value | Confidence interval | Calculated *t* value | Confidence interval |
| $t_{CSPHMM2s, LTRSPHMM1s}$ | 1.794 | [0.362, 7.438] | 1.758 | [0.220, 5.980] |
| $t_{CSPHMM2s, LTRSPHMM2s}$ | 1.756 | [0.163, 5.437] | 1.729 | [0.062, 4.339] |
| $t_{CSPHMM2s, CSPHMM1s}$ | 1.693 | [0.099, 5.701] | 1.672 | [0.042, 4.158] |

Table 8

Talking condition identification performance in stressful talking environments using the collected database based on CSPHMM2s when $\alpha = 0.5$

| Gender | Identification performance under each stressful talking condition (%) | | | | | |
|---|---|---|---|---|---|---|
| | Neutral | Shouted | Slow | Loud | Soft | Fast |
| Male | 97 | 63 | 75 | 73 | 75 | 73 |
| Female | 96 | 62 | 73 | 73 | 75 | 72 |
| Average | 96.5 | 62.5 | 74 | 73 | 75 | 72.5 |

Table 9

Emotion identification performance in emotional talking environments using the collected database based on CSPHMM2s when $\alpha = 0.5$

| Gender | Identification performance under each emotion (%) | | | | | |
| --- | --- | --- | --- | --- | --- | --- |
| | Neutral | Angry | Sad | Happy | Disgust | Fear |
| Male | 97 | 58 | 71 | 70 | 69 | 70 |
| Female | 97 | 57 | 72 | 72 | 70 | 70 |
| Average | 97 | 57.5 | 71.5 | 71 | 69.5 | 70 |

Table 10

Average stressful and emotional talking condition identification performance based on each of CSPHMM2s, SVM, GA, and VQ

| Classifier | Average stressful talking condition identification performance | Average emotional talking condition identification performance |
|---|---|---|
| CSPHMM2s | 76.3% | 73.6% |
| SVM | 72.8% | 70.9% |
| GA | 68.7% | 66.6% |
| VQ | 68.4% | 66.1% |